\documentclass{ws-spin}

\usepackage{multicol}
\usepackage{bm}
\usepackage{amsmath}
\usepackage{amssymb}
\usepackage[super]{cite}

\begin{document}

\catchline{x}{x}{2013}{}{}
\markboth{F. Mahfouzi and B. K. Nikoli\'{c}}{Nonequilibrium density matrix for spin-transfer torque calculations}

\title{HOW TO CONSTRUCT THE PROPER GAUGE-INVARIANT DENSITY MATRIX IN STEADY-STATE NONEQUILIBRIUM: APPLICATIONS TO SPIN-TRANSFER AND SPIN-ORBIT TORQUES}

\author{FARZAD MAHFOUZI and BRANISLAV K. NIKOLI\'C\footnote{bnikolic@udel.edu}}
\address{Department of Physics and Astronomy, University of Delaware, Newark, DE 19716-2570, USA}

\maketitle

\begin{history}
\received{Day Month Year}
\revised{Day Month Year}
\end{history}

\begin{abstract}
Experiments observing spin density and spin currents (responsible for, e.g., spin-transfer torque) in spintronic devices measure only the nonequilibrium contributions to these quantities, typically driven by injecting unpolarized charge current or by applying external time-dependent fields. On the other hand, theoretical approaches to calculate them operate with both the nonequilibrium (carried by electrons around the Fermi surface) and the equilibrium (carried by the Fermi sea electrons) contributions. Thus, an unambiguous procedure should remove the equilibrium contributions, thereby rendering the nonequilibrium ones which are measurable and satisfy the {\em gauge-invariant} condition according to which expectation values of physical quantities should not change when electric potential everywhere is shifted by a constant amount. Using the framework of nonequilibrium Green functions, we delineate such procedure  which yields the proper gauge-invariant nonequilibrium density matrix in the linear-response and elastic transport regime for current-carrying steady state of an open quantum system connected to two macroscopic reservoirs. Its usage is illustrated by computing: ({\em i}) conventional spin-transfer torque (STT) in asymmetric F/I/F magnetic tunnel junctions (MTJs); ({\em ii}) unconventional STT in asymmetric N/I/F semi-MTJs with the strong Rashba spin-orbit coupling (SOC) at the I/F interface and injected current perpendicular to that plane; and ({\em iii}) current-driven spin density within a clean ferromagnetic Rashba spin-split two-dimensional electron gas (2DEG) which generates SO torque in laterally patterned N/F/I heterostructures when such 2DEG is located at the N/F interface and injected charge current flows parallel to the plane. We also compare our results for these three examples with those that would be obtained using improper expressions for the density matrix, which are often found in the literature but which arbitrarily mix nonequilibrium and equilibrium expectation values due to a violation of the gauge invariance.
\end{abstract}



\section{Introduction}

The stationary nonequilibrium density matrix $\hat{\rho}_{\rm neq}$  of current carrying steady states is one of the most fundamental objects of  nonequilibrium quantum statistical mechanics and quantum transport theory.\cite{Ventra2008,Dutt2011,Datta2000,Dhar2012} This is because it yields the expectation values of any single-particle observable, while its diagonal elements give directly the particle density.\cite{Brandbyge2002,Areshkin2010,Sanvito2011} For example,
\begin{equation}\label{eq:expectation}
A=\mathrm{Tr}\, [\hat{\rho}_{\rm neq} \hat{A}],
\end{equation}
makes it possible to compute charge current, spin current, and spin density in systems out of equilibrium when the corresponding operators (i.e., their matrix representation) are inserted as $\hat{A}$.

In the case of steady-state transport of non-interacting quasiparticles described using popular tight-binding Hamiltonians,\cite{Datta2000} one can obtain charge or spin currents in the linear-response regime using the nonequilibrium Green function (NEGF)-based expressions\cite{Ventra2008,Datta2000,Nikolic2006} that do not invoke $\hat{\rho}_{\rm neq}$ explicitly. However, the inclusion of atomistic details of the device through self-consistent Hamiltonians, typically obtained\cite{Brandbyge2002,Sanvito2011} from or fitted\cite{Barraza-Lopez2010} to density functional theory (DFT), requires the knowledge of equilibrium density matrix $\hat{\rho}_{\rm eq}$ to describe the charge transfer between different atomic species\cite{Areshkin2010} or $\hat{\rho}_{\rm neq}$ to describe the charge redistribution due to the current flow at finite bias voltage.\cite{Brandbyge2002,Areshkin2010,Sanvito2011} Otherwise, without computing the  charge redistribution and the corresponding self-consistent electric  potential profile across the device the current-voltage characteristics violates\cite{Christen1996,Hernandez2009a} gauge invariance, i.e., invariance with respect to the global shift of electric potential by a constant,
$V \rightarrow V + V_0$.

The explicit construction of $\hat{\rho}_{\rm neq}$ can play an important role in solving the complicated problem of steady-state transport in interacting quantum many-particle systems far from equilibrium. For example, the recent efforts\cite{Dutt2011} have constructed an effective equilibrium-like (i.e., written in the usual Boltzmann form) density matrix,  \mbox{$\hat{\rho}_{\rm neq}=\exp[-\beta(\hat{H}-\hat{Y})]$}, using device Hamiltonian $\hat{H}$ and an additional operator $\hat{Y}$ which encodes information about the finite bias voltage applied between the two attached electrodes.

One of the key issues in applying Eq.~\eqref{eq:expectation} to specific problems is to remove possible equilibrium contribution to a physical quantity of interest, if such quantity has a non-zero expectation value in the absence of bias voltage that is compatible with the time-reversal invariance. For example, spintronic systems are abundant in such situations: ({\em i}) since spin current operator is time-reversal invariant, it can have non-zero expectation values in the thermodynamic equilibrium, as highlighted\cite{Nikolic2006,Rashba2003,Sonin2010} by the case of equilibrium local currents in two-dimensional electron gases (2DEGs) with the Rashba spin-orbit coupling (SOC); ({\em ii}) the spin operator is not time-reversal invariant, so spin density can be non-zero in thermodynamic equilibrium on the proviso that time-reversal invariance is broken by internal or external magnetic fields; ({\em iii}) the so-called field-like or perpendicular  component\cite{Ralph2008,Brataas2012} of the vector of  spin-transfer torque (STT) in magnetic tunnel junctions (MTJs) has non-zero value in equilibrium.\cite{Xiao2008a,Tang2010}

Another example of equilibrium quantities that appear in the formalism, but are not measured in standard transport experiments, are circulating or diamagnetic currents which appear in a system breaking the time-reversal invariance either by applying an external magnetic field or due to the spontaneous magnetization. They contribute to the local charge current density, ${\bf j}({\bf r}) = \int d{\bf r}^\prime \, \underline{\sigma}({\bf r},{\bf r}^\prime) {\bf E} ({\bf r}^\prime)$, which is signified by dependence of the Kubo nonlocal conductivity tensor $\underline{\sigma}({\bf r},{\bf r}^\prime)$ on {\em all} states below the Fermi energy.\cite{Baranger1989} Thus, theoretical description of charge transport in multiterminal
Hall  bridges must remove diamagnetic currents in order to produce experimentally measurable quantities, such as conductance coefficients connecting voltages and total charge currents in different terminals, which depend only on the states in some shell (defined by the temperature) around the Fermi surface.\cite{Baranger1989}

Similarly, na\"{i}ve applications of the Kubo formula to the thermal Hall coefficient always yields unphysical result due to the presence of the equilibrium circulating energy flow.\cite{Qin2011} That is, when time-reversal invariance is broken by magnetic field, the temperature gradient appears to be  driving both the transport and the circulating heat currents. Although both contributions are present in the microscopic current density calculated by the standard linear-response theory, a proper subtraction of circulating component is necessary since such quantity is {\em not observable} in the transport experiments.

On the other hand, in the literature on quantum transport of spin and charge one often finds expressions\cite{Haney2007,Heiliger2008b,Reynoso2006,Xing2010} for $\hat{\rho}_{\rm neq}$ which yield ambiguous results for the nonequilibrium expectation value of quantities like spin density, STT and local spin or charge currents. That is, being gauge non-invariant such expressions improperly subtract expectation value in equilibrium (computed at zero bias voltage), where ambiguity arises due to dependence on the chosen way of splitting the bias voltage between the source and the drain electrodes of the device.

\begin{figure*}\label{fig:fig1}
\begin{center}
\psfig{file=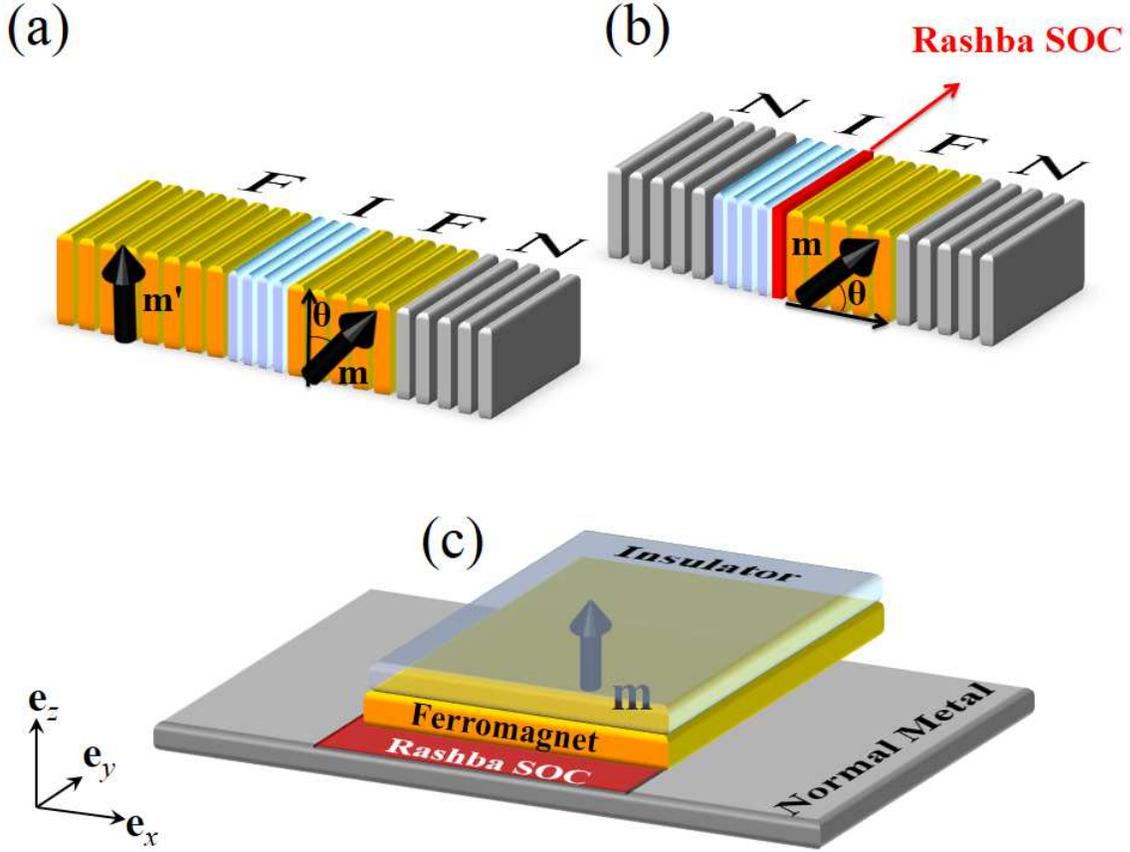,width=6in,angle=0}
\end{center}
\caption{Schematic view of junctions exhibiting different types of spin torque which are employed in Sec.~\ref{sec:applications}
to illustrate usage of the gauge-invariant nonequilibrium density matrix derived in Sec.~\ref{sec:rho}:
(a) conventional F$^\prime$/I/F MTJ consisting of a semi-infinite spin-polarizing F$^\prime$ layer with fixed magnetization and F layer of finite thickness
with free magnetization which are separated by a thin insulating I layer; (b) N/I/F semi-MTJ containing a single F layer of finite thickness with free
magnetization where the strong Rashba SOC is assumed to be located\cite{Gmitra2013} at the I/F interface; and (c) laterally patterned N/F/I heterostructure (realized in recent experiments\cite{Miron2010,Miron2011,Gambardella2011,Liu2011b,Liu2012,Liu2012c,Suzuki2011} as, e.g., Pt/Co/AlO$_x$ or Ta/CoFeB/MgO multilayers) where the strong Rashba SOC is assumed to be located\cite{Park2013} at the N/F interface. In all three cases, the unpolarized charge current driving spin torque is injected along the $x$-axis. We assume that each layer in panels (a) and (b) is composed of atomic monolayers modeled on an infinite square tight-binding lattice. The N/F interface in panel (c) is infinite in the $y$-direction and finite in the $x$-direction.}
\end{figure*}

The STT is a phenomenon in which a spin current of sufficiently large density injected into a ferromagnetic (F) layer either switches its magnetization from one static configuration to another or generates a dynamical situation with steady-state precessing magnetization.\cite{Ralph2008,Brataas2012} The origin of STT is the absorption of the itinerant flow of angular momentum components normal to the magnetization direction. It represents one of the central phenomena of the second-generation spintronics, focused on manipulation of coherent spin states, since the reduction of current densities (currently of the order 10$^6$-10$^8$ A/cm$^2$) required for STT-based magnetization switching is expected to bring about commercially viable magnetic random access memories.\cite{Katine2008} The rich nonequilibrium physics\cite{Wang2011} arising in the interplay of spin currents carried by fast conduction electrons and slow collective magnetization dynamics is also of great fundamental interest.

For conventional F$^\prime$/I/F magnetic tunnel junctions (MTJs) illustrated in Fig.~1(a), where the reference F$^\prime$ layer with fixed magnetization ${\bf m}^\prime$ plays the role of an external spin-polarizer and I stands for a thin insulating barrier, it is customary to analyze the in-plane (also called ``damping-like''\cite{Slonczewski1996})  and perpendicular  (also called ``field-like''\cite{Ralph2008,Brataas2012}) components of the STT vector
\begin{equation}\label{eq:sttcomponents}
{\bf T} = {\bf T}_{\parallel} + {\bf T}_{\perp}.
\end{equation}
The in-plane torque
\begin{equation}\label{eq:parallel}
{\bf T}_{\parallel}=\tau_{\parallel}{\bf m} \times ({\bf m} \times {\bf m}^\prime),
\end{equation}
is purely nonequilibrium and competes with the damping. The perpendicular torque
\begin{equation}\label{eq:perp}
{\bf T}_{\perp}=\tau_{\perp} {\bf m} \times {\bf m}^\prime,
\end{equation}
arises from spin reorientation at the interfaces and possesses both the equilibrium (i.e., interlayer exchange coupling) and the nonequilibrium contributions which act like an effective magnetic field on the magnetization ${\bf m}$ of the free F layer.

The often presumed\cite{Manchon2011b} bias voltage dependence of STT components in MTJs, \mbox{$\tau_{\parallel}=a_1 V_b + a_2 V_b^2$} and \mbox{$\tau_{\perp}=b_0 + b_1 V_b^2$}, is violated in asymmetrically designed MTJs (where the bias dependence of $\tau_{\perp}$ acquires a linear $\propto V_b$ contribution\cite{Oh2009}) or at large $V_b$ where recent experiments\cite{Wang2011} have uncovered deviations from these simple formulas. The most accurate STT experiments (such as those based on spin-transfer-driven ferromagnetic resonance\cite{Wang2011}) have access only to the derivatives of ${\bf T}_{\parallel}$ and ${\bf T}_{\perp}$ with respect to $V_b$. Thus, only the nonequilibrium contributions driven by the nonzero bias voltage \mbox{$V_b = V_L - V_R$} between the left and the right electrode are accessed experimentally. Accordingly, the equilibrium contribution\cite{Xiao2008a,Tang2010} to ${\bf T}_{\perp}$ (such as $b_0$ above) should be removed when comparing theoretical predictions with experimental results.

At finite bias voltage, one can simply compute \mbox{${\bf T}_\perp(V_b)-{\bf T}_\perp(V_b=0)$} to extract numerically the purely nonequilibrium perpendicular torque.\cite{Tang2010,Theodonis2006} However, this fails at small $V_b$ due to substantial numerical errors accumulated when subtracting two nearly equal numbers. Also, in the spirit of the linear-response theory, one should be able to express near-equilibrium transport quantities solely in terms of wavefunctions or GFs computed at zero bias voltage.\cite{Baranger1989} In symmetric MTJs with identical F and F$^\prime$ layers, one can eliminate equilibrium value of ${\bf T}_{\perp}$ by using the special gauge (i.e., the reference level for the electric potential) where voltage $-V_b/2$ is applied to the left and $V_b/2$ to the right electrode, or vice versa.\cite{Xiao2008a} This trick, however, is not applicable to setups with F and F$^\prime$ layers of different thickness or when they are made of different materials.\cite{Oh2009,Heiliger2008}

Moreover, it cannot be applied to recently predicted\cite{Manchon2011,Mahfouzi2012a} unconventional STT driven by SOC in asymmetric N/I/F or N/TI/F {\em vertical}  heterostructures, where TI is a thin slab of three-dimensional topological insulator\cite{Hasan2010} and strong Rashba SOC exists at I/F or TI/F interface, as illustrated in Fig.~1(b). In the case of such ``semi-MTJs'' containing only one F layer, current flowing perpendicularly through the I/F or TI/F interface will induce both ${\bf T}_{\perp}$ and ${\bf T}_{\parallel}$ on the free F layer even in the linear-response regime. This happens in the absence of any additional spin-polarizing F$^\prime$ layer through a mechanism closely related to the tunneling anisotropic magnetoresistance\cite{Matos-Abiague2009,Mahfouzi2012} (TAMR) in the case of N/I/F junctions,\cite{Manchon2011,Mahfouzi2012a} or a combination of TAMR-based effect and spin-polarizing action of the TI slab (where current becomes polarized in the direction of transport)  in the case of N/TI/F junctions.\cite{Mahfouzi2012a}

Another type of unconventional torque driven by SOC has attracted considerable attention recently.\cite{Miron2010,Miron2011,Gambardella2011,Liu2011b,Liu2012,Liu2012c,Suzuki2011,Fan2013} The so-called SO torques occur in laterally patterned N/F/I (or N/F) heterostructures composed of a single ultrathin (typically thinner than $<1$ nm) F layer in contact with N layer made of a heavy metal with large atomic SOC. The structural inversion asymmetry of these heterostructures can generate strong interfacial Rashba SOC, as illustrated in Fig.~1(c) where we assume (as suggested by the recent first-principles analysis\cite{Park2013}) that SOC is located at the N/F interface.

It has been known for a long time,\cite{Edelstein1990,Inoue2003} and confirmed by the recent experiments on semiconductor heterostructures,\cite{Ganichev2006a}  that in-plane longitudinal dc charge current flowing through a Rashba spin-split 2DEG will induce nonequilibrium spin density pointing in the transverse direction. This mechanism---the so-called Edelstein effect---provides one of the two possible explanations\cite{Manchon2012,Manchon2008,Manchon2009,Garate2009,
Obata2008,Matos-Abiague2009a,Linder2013,Wang2012,Pesin2012a,Haney2013,Pauyac2013} for the very recent experimental observations\cite{Miron2010,Miron2011,Gambardella2011,Liu2011b,Liu2012,Liu2012c,Suzuki2011,Fan2013} of magnetization switching of Co (or CoFeB or Py) layer within Pt/Co/AlO$_x$ (or Ta/CoFeB/MgO or Pt/Py) heterostructures with structural inversion asymmetry where current flows parallel to the N/F plane. The key quantity that has to be calculated in the theoretical analysis\cite{Wang2012} of SO torques is the nonequilibrium spin density within the Rashba 2DEG with an additional Zeeman term generated by the proximity to the F layer. Thus, the usage of gauge-noninvariant $\hat{\rho}_{\rm neq}$ expressions\cite{Haney2007,Heiliger2008b,Reynoso2006} here  would give an incorrect result due to the fact that this system has non-zero spin density in equilibrium, which is allowed by the time-reversal symmetry breaking due to the Zeeman term.

We provide a detailed explanation in Sec.~\ref{sec:rho} on how to construct the proper gauge-invariant expression for the nonequilibrium density matrix $\hat{\rho}_{\rm neq}$ while using widely employed NEGFs\cite{Ventra2008,Datta2000,Stefanucci2013} for devices attached to two macroscopic reservoirs at different electrochemical potentials. Our key results---Eq.~\eqref{eq:rhoneqfull} and its zero-temperature version Eq.~\eqref{eq:rhoneqfullzero}---ensures that no equilibrium contribution is included in the current-induced nonequilibrium expectation values of physical quantities. Section~\ref{sec:applications} shows  applications of this formalism to the computation of the components of conventional (in F/I/F MTJs) and unconventional (i.e., driven by the interfacial SOC in vertical N/I/F or laterally patterned N/F/I heterostructures) spin torque for junctions illustrated in Fig.~1. For each of the three cases, we also show the size of the error generated by the usage of popular but improper gauge-noninvariant expressions\cite{Haney2007,Heiliger2008b,Reynoso2006,Xing2010} for $\hat{\rho}_{\rm neq}$. We conclude in Sec.~\ref{sec:conclusions}.

\section{Gauge-invariant nonequilibrium density matrix for steady-state transport in the linear-response and elastic regime}\label{sec:rho}

Let us consider a finite-size open quantum system described on a tight-binding lattice\cite{Datta2000,Nikolic2006} where the operator $\hat{c}_{n \sigma}^\dag$ ($\hat{c}_{n \sigma}$) creates (annihilates) electron with spin $\sigma$ on site $n$ (specific examples of such Hamiltonians are given in Sec.~\ref{sec:applications}).  The NEGF formalism\cite{Stefanucci2013} operates with two fundamental objects---the retarded
\mbox{$G^{r,\sigma\sigma'}_{nn'}(t,t')=-i \Theta(t-t') \langle \{\hat{c}_{n\sigma}(t) , \hat{c}^\dagger_{n'\sigma'}(t')\}\rangle$} and the lesser
\mbox{$G^{<,\sigma\sigma'}_{nn'}(t,t')=i \langle \hat{c}^\dagger_{n'\sigma'}(t') \hat{c}_{n \sigma}(t)\rangle$} GF that describe the density of available quantum states and how electrons occupy those states, respectively. Here $\langle \ldots \rangle$ denotes the nonequilibrium statistical average.\cite{Stefanucci2013} In stationary problems, $\hat{G}^r$ and $\hat{G}^<$ depend only on the time difference $t-t^\prime$, or energy $E$ after
the Fourier transform.\cite{Nikolic2006}

We assume that the system is opened by being attached to two macroscopic reservoirs---left (L) and right (R)---which drive charge current when they have different electrochemical potentials and, thereby, different  Fermi functions \mbox{$f_{L,R}(E)=f(E-eV_{L,R})$}. The reservoirs and dissipation they are responsible
for do not have to be modeled explicitly. Instead, one introduces the left and the right semi-infinite ideal leads through their retarded self-energies\cite{Velev2004} $\hat{\Sigma}_{L,R}^r(E)$, so that Hamiltonian \mbox{$\hat{H} + \hat{\Sigma}_{L}^r(E) + \hat{\Sigma}_{R}^r(E)$} of a finite-size but open quantum system acquires a continuous spectrum. The continuous spectrum  is sufficient to bring the system into a true nonequilibrium
steady-state with a finite value of dc current at long enough times as demonstrated by, e.g.,  real-time diagrammatic Monte Carlo simulations of nonequilibrium quantum transport\cite{Schiro2009}.

In stationary situations, either due to thermodynamic equilibrium or steady-state current flow, the density matrix $\hat{\rho}$ can be expressed\cite{Stefanucci2013} in terms of the lesser GF
\begin{equation}\label{eq:rhoglesser}
\hat{\rho} = \frac{1}{2\pi i} \int dE\, \hat{G}^<(E).
\end{equation}
In the case of elastic transport regime (i.e., when electron-electron, electron-phonon, and electron-spin dephasing processes can be neglected), the lesser GF
\begin{equation}\label{eq:keldysh}
\hat{G}^<(E)=\hat{G}^r(E) \left[i f_L(E) \hat{\Gamma}_L(E) +  i f_R(E) \hat{\Gamma}_R(E) \right] \hat{G}^a(E),
\end{equation}
is given solely in terms of the retarded GF, $\hat{G}^r(E)$. Here $\hat{G}^a(E) = [\hat{G}^r(E)]^\dagger$ is the advanced GF, and $\hat{\Gamma}_{L,R}(E)=i[\hat{\Sigma}_{L,R}^r(E)-\hat{\Sigma}_{L,R}^a(E)]$ are the level broadening operators which quantify escape rates of
electrons into the semi-infinite leads. The usual assumption about the leads is that the applied bias voltage $V_b$ induces a rigid shift
in their electronic structure,\cite{Brandbyge2002} so that
\begin{eqnarray}\label{eq:rigid}
\hat{\Sigma}_{L,R}(E,V_b) & = & \hat{\Sigma}_{L,R}(E-eV_{L,R}), \\
\hat{\Gamma}_{L,R}(E,V_b) & = & \hat{\Gamma}_{L,R}(E-eV_{L,R}).
\end{eqnarray}

To simplify computation, integration in Eq.~\eqref{eq:rhoglesser} for the elastic transport regime
version of $\hat{G}^<(E)$ in Eq.~\eqref{eq:keldysh} is typically separated into apparent ``equilibrium'' and ``nonequilibrium'' terms as follows.\cite{Brandbyge2002,Areshkin2010} We first add and subtract the term $\hat{G}^r \hat{\Gamma}_L \hat{G}^a f_R(E)$ to Eq.~\eqref{eq:keldysh},
to get after some rearrangements
\begin{eqnarray}\label{eq:rearrange}
\hat{G}^<(E)  =  i \hat{G}^r (\hat{\Gamma}_L + \hat{\Gamma}_R) \hat{G}^a f_R(E) + i \hat{G}^r \hat{\Gamma}_L \hat{G}^a [f_L(E) - f_R(E)].
\end{eqnarray}
By substituting the following identity
\begin{eqnarray}\label{eq:identity}
\hat{\Gamma}_L + \hat{\Gamma}_R = i \left[ (\hat{G}^a)^{-1} - (\hat{G}^r)^{-1} \right],
\end{eqnarray}
into Eq.~\eqref{eq:rearrange}, we finally obtain
\begin{equation}\label{eq:im}
\hat{G}^<(E)  =  i (\hat{G}^r - \hat{G}^a) f_R(E) + i \hat{G}^r \hat{\Gamma}_L \hat{G}^a [f_L(E) - f_R(E)].
\end{equation}
This allows us to rewrite Eq.~\eqref{eq:rhoglesser} as the sum of two contributions
\begin{equation}\label{eq:rho}
\hat{\rho}   =  -\frac{1}{\pi} \int\limits_{-\infty}^{+\infty} dE \, {\rm Im}\left[ \hat{G}^r(E) \right] f(E-eV_R) +  \frac{1}{2 \pi} \int\limits_{-\infty}^{+\infty}dE \,  \hat{G}^r(E) \cdot \hat{\Gamma}_L(E-eV_L) \cdot   \hat{G}^a(E) \left[ f(E-eV_L) - f(E-eV_R) \right].
\end{equation}
The first ``equilibrium'' term contains integrand which is analytic in the upper complex plane, so that it can be computed via the semicircular path combined with the path in the upper complex plane parallel to the real axis.\cite{Brandbyge2002,Areshkin2010} Because the functions $\hat{G}^r(E)$ and $\hat{G}^a(E)$ are nonanalytic below and above the real axis, respectively, the integrand in the second ``nonequilibrium'' term is nonanalytic function in the entire complex energy plane, so that integration\cite{Areshkin2010,Sanvito2011} has to be done directly along the real axis between the boundaries around $E_F-eV_R$ and $E_F-eV_L$ determined by the difference of the Fermi functions ($E_F$ is the Fermi energy for the whole device in equilibrium).

While the second ``nonequilibrium'' term in Eq.~\eqref{eq:rho} contains information about the bias voltage [through the difference \mbox{$f_L(E)-f_R(E)$}], as well as about the lead assumed to be injecting electrons into the device (through $\hat{\Gamma}_L$), it cannot be used as the proper nonequilibrium density matrix which is defined by
\begin{equation}\label{eq:rhoneq}
\hat{\rho}_{\rm neq} = \hat{\rho} -\hat{\rho}_{\rm eq} = \hat{\rho} + \frac{1}{\pi} \int\limits_{-\infty}^{+\infty} dE \, {\rm Im}\, \left[\hat{G}_0^r(E)\right] f(E),
\end{equation}
where
\begin{equation}
\hat{G}^r_0(E) = \left[ E- \hat{H} - \hat{\Sigma}_L(E) - \hat{\Sigma}_R(E) \right]^{-1},
\end{equation}
is the retarded GF at zero bias voltage. This is due to the fact that second term in Eq.~\eqref{eq:rhoneq}, as the NEGF expression for the equilibrium density matrix $\hat{\rho}_{\rm eq}$, does not cancel the gauge-noninvariant first term in Eq.~\eqref{eq:rho} which depends explicitly [through $f(E-eV_R)$] on the arbitrarily set $V_R$ and implicitly on the voltages applied to both reservoirs [through $\hat{G}^r(E)$]. Nevertheless, the second term in Eq.~\eqref{eq:rho},
\begin{eqnarray}\label{eq:incorrectfull}
 \frac{1}{2 \pi} \int\limits_{-\infty}^{+\infty}dE \,  \hat{G}^r(E) \cdot \hat{\Gamma}_L(E-eV_L) \cdot \hat{G}^a(E) [f_L(E)-f_R(E)],
\end{eqnarray}
written in the linear-response and typically zero-temperature limit (where it becomes the Fermi surface property)
\begin{equation} \label{eq:incorrect}
\frac{eV_b}{2 \pi} \hat{G}^r_0(E_F) \cdot \hat{\Gamma}_L(E_F) \cdot \hat{G}^a_0(E_F),
\end{equation}
is often used in the quantum transport literature\cite{Haney2007,Heiliger2008b,Reynoso2006,Xing2010} as the putative but improper (due to being gauge-noninvariant)  expression for $\hat{\rho}_{\rm neq}$. Thus, its usage leads to ambiguous (i.e., dependent on the chosen $V_R$) nonequilibrium
expectation values.

To derive the proper gauge-invariant $\hat{\rho}_{\rm neq}$ in the linear-response limit, we first expand the retarded GF
\begin{eqnarray}\label{eq:gr}
\hat{G}^r(E)   =   \left[ E- \hat{H} - eU - \hat{\Sigma}_L(E-eV_L) - \hat{\Sigma}_R(E-eV_R) \right]^{-1},
\end{eqnarray}
to linear order in the bias voltage $V_b$. Here $eU$ is the potential profile across the active region of the device when the current is flowing which interpolates between $V_L$ and $V_R$. This is achieved in two steps, where we first rewrite Eq.~\eqref{eq:gr} using the exact Dyson equation\cite{Hernandez2009a}
\begin{eqnarray}\label{eq:dyson}
\hat{G}^r(E)  =   \hat{G}_0^r(E) + \hat{G}_0^r(E) \left[eU + \hat{\Sigma}_L(E-eV_L)  - \hat{\Sigma}_L(E) + \hat{\Sigma}_R(E-eV_R) - \hat{\Sigma}_R(E) \right] \hat{G}^r(E).
\end{eqnarray}
In the second step, we expand the self-energies
\begin{equation}\label{eq:sigma}
\hat{\Sigma}_{L,R}(E-eV_L) \approx \hat{\Sigma}_{L,R}(E) - eV_{L,R} \frac{\partial \hat{\Sigma}_{L,R}}{\partial E}\biggr|_{V_{L,R}=0},
\end{equation}
to linear order in voltage. Combining Eqs.~\eqref{eq:dyson} and \eqref{eq:sigma} gives
\begin{eqnarray}\label{eq:lineargr}
\hat{G}^r(E)  \approx  \hat{G}_0^r(E) + \hat{G}_0^r(E) \left[eU - eV_L \frac{\partial \hat{\Sigma}_L}{\partial E}\biggr|_{V_{L}=0} - eV_R \frac{\partial \hat{\Sigma}_R}{\partial E}\biggr|_{V_{R}=0} \right] \hat{G}^r_0(E).
\end{eqnarray}

By plugging Eq.~\eqref{eq:lineargr} into Eq.~\eqref{eq:rhoneq}, together with the expansion of the Fermi functions
\begin{equation}\label{eq:fermi}
f_{L,R}(E) \approx f(E) - eV_{L,R} \frac{\partial f}{\partial E}\biggr|_{V_{L,R}=0},
\end{equation}
and the analogous expansion of the level broadening operator
\begin{equation}\label{eq:gamma}
\hat{\Gamma}_{L}(E-eV_L) \approx \hat{\Gamma}_{L}(E) - eV_{L} \frac{\partial \hat{\Gamma}_{L}}{\partial E}\biggr|_{V_{L}=0},
\end{equation}
and by keeping only the terms linear in the applied voltage, we finally obtain the gauge-invariant nonequilibrium density matrix for the steady-state transport in the linear-response and elastic regime
\begin{eqnarray}\label{eq:rhoneqfull}
\hat{\rho}_{\rm neq}  & = & -\frac{eV_R}{\pi} \int\limits_{-\infty}^{+\infty} \!\!dE \, \mathrm{Im} \left[G_0^r\right] \left(-\frac{\partial f}{\partial E}\right) - \frac{1}{\pi} \int\limits_{-\infty}^{+\infty} \!\! dE \, \mathrm{Im} \left[ \hat{G}^r_0 \left(eU - eV_L \frac{\partial \hat{\Sigma}_L}{\partial E} - eV_R \frac{\partial \hat{\Sigma}_R}{\partial E} \right) \hat{G}^r_0 \right]f(E) \nonumber \\  
&& \mbox{} + \frac{eV_b}{2\pi} \int\limits_{-\infty}^{+\infty} \!\! dE \, \hat{G}^r_0 \hat{\Gamma}_L \hat{G}^a_0 \left(-\frac{\partial f}{\partial E}\right).
\end{eqnarray}
In the zero-temperature limit, this expression simplifies to
\begin{eqnarray}\label{eq:rhoneqfullzero}
 \hat{\rho}_{\rm neq} & = & - \frac{eV_R}{\pi} \mathrm{Im}\left[G_0^r(E_F)\right] - \frac{1}{\pi} \int\limits_{-\infty}^{E_F}dE \, \mathrm{Im} \left[ \hat{G}^r_0 \left(eU - eV_L \frac{\partial \hat{\Sigma}_L}{\partial E} - eV_R \frac{\partial \hat{\Sigma}_R}{\partial E} \right) \hat{G}^r_0 \right] f(E) \nonumber \\
 && \mbox{} + \frac{eV_b}{2\pi} \hat{G}^r_0 (E_F) \cdot \hat{\Gamma}_L(E_F) \cdot \hat{G}^a_0(E_F).
\end{eqnarray}
We note that expansions discussed above could be performed further\cite{Hernandez2009a} to obtain $\hat{\rho}_{\rm neq}$  order-by-order in the applied bias voltage.

The third term in Eq.~\eqref{eq:rhoneqfull}, when traced with the total current operator\cite{Datta2000} $\hat{I} = 2e\hat{\Gamma}_R/\hbar$ in the right lead, gives the usual Landauer-type conductance formula\cite{Caroli1971}
\begin{eqnarray} \label{eq:caroli}
G  =  \frac{I}{V_b} = \mathrm{Tr}[\hat{\rho}_{\rm neq} \hat{I}] = \frac{2e^2}{h} \int\limits_{-\infty}^{+\infty} \!\! dE \, \mathrm{Tr}\left[\hat{\Gamma}_R \hat{G}^r_0 \hat{\Gamma}_L \hat{G}^a_0 \right] \left(-\frac{\partial f}{\partial E}\right).
\end{eqnarray}
The same trace with the first two terms in Eq.~\eqref{eq:rhoneqfull} is identically equal to zero because no total charge current can flow into the leads in thermodynamic equilibrium, even if time-reversal invariance is broken by magnetic field.\cite{Baranger1989}

The first and second term in Eq.~\eqref{eq:rhoneqfullzero} make this expression for $\hat{\rho}_\mathrm{neq}$ quite different from Eq.~\eqref{eq:incorrect}. Their role is to properly subtract any non-zero expectation value that exists in thermodynamic equilibrium. For example, the first term in Eq.~\eqref{eq:rhoneqfullzero} is easily interpreted using $\hat{\rho}_\mathrm{eq}$ in Eq.~\eqref{eq:rhoneq}---when traced with an operator this term will give equilibrium expectation value governed by the states at the Fermi energy which must be removed [note that the sign in front of the first term is different from the sign in front of the third term in Eq.~\eqref{eq:rhoneqfullzero}]. The second term in Eq.~\eqref{eq:rhoneqfullzero} ensures the gauge invariance of the nonequilibrium
expectation values, while making the whole expression non-Fermi-surface property. It also renders the usage of  Eq.~\eqref{eq:rhoneqfullzero} computationally demanding due to the requirement to perform integration from the bottom of the band up to the Fermi energy, as discussed in more detail in Sec.~\ref{sec:applications}.

\section{Applications to spin torque calculations}~\label{sec:applications}

The NEGF formalism offers three different algorithms\cite{Haney2007,Theodonis2006,Mahfouzi2012a} to compute STT at finite bias voltage in F$^\prime$/I/F MTJs, which are delineated below together with their limits of applicability. There is also an additional algorithm\cite{Carva2009} for STT in the linear-response regime (therefore suitable for F$^\prime$/N/F spin valves) based on the GF expressions for spin-dependent interface conductances which enter generalized set of Kirchhoff laws describing transport properties and magnetization dynamics in a circuit containing noncollinear magnetic elements.\cite{Brataas2006}

All of these methodologies treat elastic transport of electrons, which can become insufficient when the bias voltage applied to MTJs exceeds few hundreds mV. Inelastic effects like electron-magnon and electron-phonon scattering on STT can also be handled by NEGF formalism, but the evaluation\cite{Reininghaus2006} of the corresponding lesser and retarded self-energies using nonequilibrium many-body perturbation theory\cite{Stefanucci2013} makes such algorithms computationally far more demanding and has been mostly avoided thus far.

The often employed\cite{Tang2010,Theodonis2006} model in calculations of elastic contribution to STT in MTJs in Fig.~1(a) is defined on a simple cubic lattice, with lattice constant $a$ and unit area $a^2$, where monolayers of different materials (F, N, I) are infinite in the transverse  direction (i.e., $yz$-planes). In the simplest case, the F, N, and I layers are described by a tight-binding Hamiltonian with a single $s$-orbital per site
\begin{eqnarray} \label{eq:fnhamiltonian}
\hat{H}_{F,I,N}  =  \sum_{n,\sigma\sigma',{\bf k}_{\parallel}} \hat{c}_{n\sigma,{\bf k}_{\parallel}}^{\dagger} \left(\varepsilon_{n,{\bf k}_{\parallel}} \delta_{\sigma\sigma'} - \frac{\Delta_n}{2} \mathbf{m} \cdot [\hat{\bm \sigma}]_{\sigma \sigma'} \right) \hat{c}_{n\sigma',{\bf k}_{\parallel}}  - \gamma \sum_{n,\sigma,{\bf k}_{\parallel}}  (\hat{c}_{n\sigma,{\bf k}_{\parallel}}^{\dagger}\hat{c}_{n+1, \sigma,{\bf k}_{\parallel}} + \mathrm{H.c.}).
\end{eqnarray}
The operator $\hat{c}_{n\sigma,{\bf k}_{\parallel}}^{\dagger}$ ($\hat{c}_{n\sigma,{\bf k}_{\parallel}}$) creates (annihilates) electron with spin $\sigma$ on monolayer $n$ with transverse momentum
${\bf k}_{\parallel}=(k_y,k_z)$ within the monolayer, and the nearest neighbor hopping is \mbox{$\gamma=1.0$ eV}. The in-monolayer kinetic energy \mbox{$\varepsilon_{2D} = -2\gamma(\cos k_y a + \cos k_z a)$} is equivalent to an increase in the on-site energy \mbox{$\varepsilon_{n,{\bf k}_{\parallel}} \mapsto \varepsilon_{n,{\bf k}_{\parallel}} + \varepsilon_{2D}$}. Here $\hat{\bm \sigma}=(\hat{\sigma}_x,\hat{\sigma}_y,\hat{\sigma}_z)$ is the vector of the Pauli matrices and $[\hat{\sigma}_\alpha]_{\sigma\sigma'}$ denotes the Pauli matrix elements. The coupling of itinerant electrons to collective magnetization is described through the material-dependent mean-field exchange splitting \mbox{$\Delta_n$}, where $\Delta_n \equiv 0$ within semi-infinite ideal N leads or within the tunnel barrier region I (the corresponding Hamiltonians are labeled by $\hat{H}_F$, $\hat{H}_N$, and  $\hat{H}_I$).

The dynamics of spin density for itinerant electrons is governed by the spin continuity equation\cite{Manchon2011b}
\begin{equation}\label{eq:eom}
\frac{\partial \mathbf{S}}{\partial t} =  \nabla \cdot \mathcal{J}^S + \frac{\mathbf{S}_\mathrm{neq}}{\tau_\mathrm{sf}} + \mathbf{T},
\end{equation}
where $\mathcal{J}^S$ is the spin current tensor and \mbox{$\mathbf{S}_\mathrm{neq} = \mathbf{S} - \mathbf{S}_\mathrm{eq}$} is the nonequilibrium spin density (responsible for STT). Here the spin relaxation time $\tau_\mathrm{sf}$ is introduced phenomenologically to model the spin-flip processes by impurities and
magnons. In the steady-state $\partial \mathbf{S} / \partial t \equiv 0$, and $\tau_\mathrm{sf} \approx 0$ when SOC and other spin relaxation mechanisms
can be neglected (diffusive spin-dependent transport and spatial variation of spin density are taken into account in spin valves F/N/F, but typically
neglected in F/I/F MTJs because their resistance and magnetoresistance are dominated by the electronic states near the I barrier). This makes it possible
to express STT simply as the divergence of spin current
\begin{equation}\label{eq:divergence}
\mathbf{T} = \nabla \cdot \mathcal{J}^S,
\end{equation}
whose discretized form
\begin{equation}\label{eq:flux}
{\bf T}_{n} = -\nabla \cdot {\bf I}^S = {\bf I}_{n-1,n}^S -  {\bf I}_{n,n+1}^S,
\end{equation}
gives the monolayer-resolved STT. The total torque on the free magnetization of the F layer is then obtained from
\begin{equation}
{\bf T} = \sum_{n=0}^\infty  ({\bf I}_{n-1,n}^S - {\bf I}_{n,n+1}^S) = {\bf I}_{-1,0}^S - {\bf I}_{\infty,\infty}^S = \mathbf{I}^S_{-1,0}.
\end{equation}
Here the subscripts -1 and 0 refer to the last monolayer of the I barrier and the first monolayer of the free F layer, respectively, and it is assumed that ${\bf I}_{\infty,\infty}^S=0$ because the components of ${\bf I}^S_{n,n+1}$ transverse to $\mathbf{m}$ decay to zero at infinity.\cite{Tang2010,Theodonis2006}

Thus, this ``net spin-current flux'' approach offers the simplest NEGF-based route\cite{Tang2010,Theodonis2006} to get STT by computing the vector of spin current between the two neighboring monolayers $n=-1$ and $n+1=0$
\begin{equation}\label{eq:localcurrent}
{\bf I}^S_{n,n+1} = \frac{\gamma}{4\pi} \int dE d\mathbf{k}_{\parallel} \, \mathrm{Tr}_\sigma \, [\bm{\sigma} (\hat{G}^<_{n+1,n} -  \hat{G}^<_{n,n+1})].
\end{equation}
The integration over $\mathbf{k}_{\parallel}$ is required because of the assumed translational invariance in the transverse direction. For F layer of finite thickness, one needs to compute the difference of such spin currents at the entrance and exit of the layer. However, since spin current approach relies on spin conservation that gives rise to Eqs.~\eqref{eq:divergence} and ~\eqref{eq:flux}, it cannot be applied to semi-MTJs in Fig.~1(b) because $\mathbf{I}^S_{-1,0}$ is {\em insufficient} to get STT if strong SOC is present directly at the I/F interface. Also, spin current will not decay, ${\bf I}^S_{\infty,\infty} \neq 0$, if SOC is present in the bulk of the free F layer.\cite{Hals2010,Haney2010}

A more general approach, which makes it possible to compute the STT vector in the presence of SOC or other spin-nonconserving interactions, is to use the torque operator\cite{Haney2007,Heiliger2008,Carva2009,Kalitsov2006}
\begin{equation}
\hat{\bf T} = \frac{d\hat{\bf S}}{dt} = \frac{1}{2i} [\hat{\bm \sigma},\hat{H}_F]
\end{equation}
and find its expectation value ${\bf T} = \mathrm{Tr}\,[\hat{\rho}_{\rm neq} \hat{\bf T}]$. Here $\hat{\mathbf{S}}=\hbar\hat{\bm \sigma}/2$ is the electron spin operator. Using $\hat{H}_F$ in Eq.~\eqref{eq:fnhamiltonian} gives $\hat{\bf T} = \Delta (\hat{\bm \sigma} \times {\bf m})/2$. The torque operator approach is applicable to finite thickness free F layers, where it gives the monolayer-resolved\cite{Haney2007,Wang2008b} STT whose sum over all monolayers comprising the free F layer gives the total STT. It also offers a microscopic picture of STT where nonequilibrium spin density of current carrying quasiparticles is misaligned with respect to spins of electrons composing the magnetic condensate.\cite{Haney2007} This causes local torque on individual atoms which can be summed to find the net effect on the order parameter of either ferromagnetic or antiferromagnetic layers.\cite{Haney2008}

The torque operator approach can also be combined with noncollinear DFT to take into account atomistic structure of the junction from first-principles.\cite{Haney2007} Unlike collinear spin-unrestricted DFT,\cite{Fiolhais2003} where the exchange-correlation (XC) functional  $E_\mathrm{XC}[n(\mathbf{r}),m_z(\mathbf{r})]$ depends solely on the total particle density \mbox{$n=n_\uparrow + n_\downarrow$} and the $z$-component of the spin density  vector \mbox{$m_z = n_\uparrow - n_\downarrow$}, noncollinear DFT functional $E_\mathrm{XC}[n(\mathbf{r}),\mathbf{m}(\mathbf{r})]$ is required when the direction of the local spin density is not constrained to a particular axis. The XC magnetic field, \mbox{${\bf B}_\mathrm{XC}(\mathbf{r}) = \delta E_\mathrm{XC}[n(\mathbf{r}),\mathbf{m}(\mathbf{r})]/\delta \mathbf{m}(\mathbf{r})$}, introduces the term $-{\bf B}_\mathrm{XC}(\mathbf{r}) \cdot \hat{\bm \sigma}$ into the single-particle self-consistent Kohn-Sham Hamiltonian, which replaces the tight-binding Hamiltonian in Eq.~\eqref{eq:tbh} within the NEGF-DFT framework.\cite{Brandbyge2002,Areshkin2010} The torque can then be computed in the coordinate representation as
\begin{equation}\label{eq:sttreapspace}
\mathbf{T} = \int\limits_F \! d^3 r \, \mathbf{m}_\mathrm{neq}(\mathbf{r}) \times \mathbf{B}_\mathrm{XC}(\mathbf{r}),
\end{equation}
where integration is performed over the volume of the free F layer and the nonequilibrium spin density is obtained from $\mathbf{m}_\mathrm{neq} = \mathrm{Tr}\, [ \hat{\rho}_\mathrm{neq} \hat{\bm \sigma} ]$.

However, most of presently implemented versions of noncollinear XC functionals have vanishing local torque $\mathbf{m}_\mathrm{eq}(\mathbf{r}) \times \mathbf{B}_\mathrm{XC}(\mathbf{r}) \equiv 0$ in equilibrium,\cite{Capelle2001,Scalmani2012} so that third term in Eq.~\eqref{eq:rhoneqfull} is sufficient to obtain current-driven torque in such approximations.\cite{Haney2007,Haney2008} Nevertheless, combining the very recently proposed\cite{Scalmani2012} noncollinear DFT having proper invariance and local torque properties with NEGF will require to properly remove the non-zero torque in equilibrium \mbox{$\int\limits_F \! d^3 r \, \mathbf{m}_\mathrm{eq}(\mathbf{r}) \times \mathbf{B}_\mathrm{XC}(\mathbf{r}) \neq 0$} by using all three terms in Eq.~\eqref{eq:rhoneqfull}.

The third\cite{Mahfouzi2012a} NEGF-based approach to the computation of STT vector is also applicable to F layers of finite thickness with bulk or interfacial SOC  while offering additional insights. If the device Hamiltonian depends on a variable $q$, which corresponds to slow collective classical degrees of freedom, the expectation value of the corresponding canonical force
\mbox{$\hat{Q} = -\partial \hat{H}/\partial q$} can be obtained from
\begin{eqnarray}\label{eq:nebo}
Q =  -\frac{1}{2\pi i} \int\limits_{-\infty}^{+\infty} dE\, \mathrm{Tr}\, \left[\frac{\partial \hat{H}}{\partial q}\hat{G}^{<} \right] =  -\left \langle\frac{\partial \hat{H}}{\partial q}\hat{G}^{<} \right \rangle.
\end{eqnarray}
Here we used the density matrix Eq.~\eqref{eq:rhoglesser} expressed in terms of $\hat{G}^{<}(E,q)$ as adiabatic lesser GF computed for a frozen-in-time variable $q$. By exchanging the derivative between
the Hamiltonian and $\hat{G}^{<}(E,q)$,
\begin{equation}
Q =-\frac{\partial \langle\hat{H}\hat{G}^{<}\rangle}{\partial q} + \left \langle \frac{\hat{H} \partial \hat{G}^{<}}{\partial q} \right \rangle,
\end{equation}
and by using Eqs.~\eqref{eq:keldysh} and \eqref{eq:gr} for the retarded and lesser GFs, respectively, we obtain
\begin{equation} \label{eq:central}
Q = i \left \langle\frac{\partial \hat{G}^r}{\partial q}\hat{\Sigma}^{<}\hat{G}^a\hat{\Gamma}\right \rangle - \left \langle \hat{\Sigma}^{<} \frac{\partial \hat{G}^r}{\partial q} \right \rangle.
\end{equation}
In the elastic transport regime, the lesser self-energy is given by\cite{Stefanucci2013} \mbox{$\hat{\Sigma}^<(E) = \sum_p i f_p(E) \hat{\Gamma}_p(E-eV_p)$} for leads $p=L,R$.

We note that Eq.~\eqref{eq:central} is akin to the mean value of time-averaged force in nonequilibrium Born-Oppenheimer approaches\cite{Bode2011,Bode2012} to current-induced forces exerted by conduction electrons on ions in nanojunctions or mechanical degrees of freedom in nanoelectromechanical systems whose collective modes are slow compared to electronic time scales. Furthermore, the same derivation that leads to Eq.~\eqref{eq:central} can be extended to obtain\cite{Bode2012} the noise and damping terms, expressed solely in terms of electronic NEGFs, which enter into the nonequilibrium Langevin equation (taking the form of a generalized Landau-Lifshitz-Gilbert equation) for the free magnetization of the F layer.

The application of Eq.~(\ref{eq:central}) to get $T_\alpha$ ($\alpha=x,y,z$) component of the STT vector acting on the magnetization of the free F layer within, e.g., F$^\prime$/I/F MTJ proceeds by first computing $\hat{G}^r$ for the device described by the Hamiltonian $\hat{H} = \hat{H}_{F^\prime} + \hat{H}_I + \hat{H}_F$. In the second step, the Hamiltonian of the F layer is modified
\begin{equation}\label{eq:hfprime}
\hat{H}_{F}^q = \hat{H}_{F} + q\sum_{n,\sigma\sigma',{\bf k}_{\parallel}}  \hat{c}_{n\sigma,{\bf k}_{\parallel}}^{\dagger} [{\bf e}_\alpha  \cdot ({\bf m} \times \hat{\bm \sigma})]_{\sigma\sigma'} \hat{c}_{n\sigma',{\bf k}_{\parallel}},
\end{equation}
and $\hat{G}^r[\hat{H}^q]$ is computed for the new Hamiltonian $\hat{H}^q  = \hat{H}_{F^\prime} + \hat{H}_I + \hat{H}_{F}^q$. This yields
\begin{equation}
\frac{\partial \hat{G}^r}{\partial q} \approx \frac{\hat{G}^r[\hat{H}^q]-\hat{G}^r[\hat{H}]}{q},
\end{equation}
where we typically employ $q = 10^{-7}$ as the infinitesimal. The derivative $\partial \hat{G}^r/\partial q$ plugged into Eq.~\eqref{eq:central} yields $Q = T_\alpha$.

Equation~(\ref{eq:central}) includes both the equilibrium\cite{Xiao2008a,Tang2010,Haney2007} \mbox{${\bf T}_\perp(V_b=0)$} and experimentally measured\cite{Wang2011} nonequilibrium \mbox{${\bf T}_\perp(V_b)-{\bf T}_\perp(V_b=0)$} contribution to ${\bf T}_\perp$. The linear-response contribution at zero temperature can be extracted by using $\hat{\rho}_{\rm neq}$ in Eq.~\eqref{eq:rhoneqfullzero}
\begin{eqnarray}\label{eq:qneq}
Q_{\rm neq}  =  -\sum_{p} V_p {\rm Tr} \left[ \frac{\partial \hat{G}^r_0}{\partial q} \hat{\Gamma}_{p} \hat{G}^a_0 \hat{\Gamma} - i\frac{\partial \hat{G}^r_0}{\partial q} \hat{\Gamma}_p \right] -\sum_{p} V_p {\rm Im} \left\{ \int\limits_{-\infty}^{E_F} \!\! dE\, {\rm Tr} \, \left[\frac{\partial \hat{G}^{r}_0}{\partial q} \frac{\partial \hat{H}}{\partial V_{p}}-\frac{\partial \hat{G}^{r}_0}{\partial q}\frac{\partial \hat{\Sigma}^r_p}{\partial E} \right] \right\}.
\end{eqnarray}
Since ${\bf T}_\parallel$ is zero in equilibrium, the second sum in Eq.~\eqref{eq:rhoneq} has to be computed only for ${\bf T}_\perp$.

Note that in computation of ${\bf T}_\parallel$, Eq.~\eqref{eq:qneq} is more efficient than the torque operator approach discussed above  since the former requires to know only the submatrix of the retarded GF which couples the first and last monolayer of the active device region, while the later requires to obtain the retarded GF on each monolayer of the free F layer. When computing ${\bf T}_\perp$, both methods have similar computational complexity since they require knowledge of the full retarded GF matrix.

When evaluating Eq.~\eqref{eq:qneq}, we use the fact that the integrand in the second term  [which stems from the second term in Eq.~\eqref{eq:rhoneqfullzero}] is analytic function in the upper complex plane. Then the integration can proceed along the contour composed of an infinite semi-circle, along which the trace is zero due to $\partial \hat{G}^r/\partial q \sim 1/E^2$, and the vertical line at $E=E_F$. We note that adaptive integration is required very close to $E_F$. For junctions with transverse translational symmetry, such as the ones in Fig.~1(a),(b), one has to perform additional integration over ${\bf k}_{\parallel}$. This requires adaptive scheme (or very dense $k$-point sampling in brute force schemes\cite{Wang2008b}) to converge the integrand because of the fact that STT can change fast in the specific regions of the 2D Brillouin zone.

Aside from our arguments based on the usage of proper gauge-invariant nonequilibrium density matrix in Eq.~\eqref{eq:rhoneqfullzero}, recent analysis\cite{Haney2009} of interlayer exchange coupling in junctions brought out of equilibrium by the applied bias voltage has also found it necessary to perform integration over the whole Fermi sea in order to obtain the correct value of ${\bf T}_\perp$ torque component.

\subsection{Application to STT in MTJs}\label{sec:mtj}

The MTJ in Fig.~1(a) is modeled by the Hamiltonian in Eq.~\eqref{eq:fnhamiltonian}, where the I layer has thickness $d_{\rm I}=5$ monolayers, $d_{\rm F}=20$ is the thickness of the free F layer and the fixed F$^\prime$ layer is semi-infinite. The $\mathrm{TMR}=(R_{AP}-R_{P})/R_{AP}$ of this MTJ is 100\%, where $R_P$ is resistance for parallel configuration of the magnetizations $\mathbf{m}$ and $\mathbf{m}^\prime$ and $R_{AP}$ is resistance for their antiparallel orientation. The I layer has $\varepsilon_n = 6.0$ eV to model the potential barrier. The Fermi energy of this device in equilibrium is set at \mbox{$E_F=3.1$} eV.  The F layers have the same mean-field exchange splitting $\Delta=1.0$ eV, but since they are not of the same thickness MTJ is asymmetric and its linear-response ${\bf T}_\perp \propto V_b$ is non-zero.\cite{Xiao2008a,Tang2010,Wang2011,Oh2009,Heiliger2008} This is confirmed in Fig.~2(a) using the proper gauge-invariant nonequilibrium density matrix in Eq.~\eqref{eq:rhoneqfullzero}. On the other hand, using the improper gauge-noninvariant expression Eq.~\eqref{eq:incorrect} for $\hat{\rho}_{\rm neq}$ gives  ${\bf T}_\perp  \propto V_b$ in Fig.~2(b) which is about two orders of magnitude smaller. Since ${\bf T}_\parallel$ does not have non-zero expectation value in equilibrium, both the proper and improper expressions for $\hat{\rho}_{\rm neq}$ give the same result. Note that Fig.~2 reproduces the well-known $\propto \sin \theta$ angular dependence for both torque components.

\begin{figurehere}\label{fig:fig2}
\centerline{\psfig{file=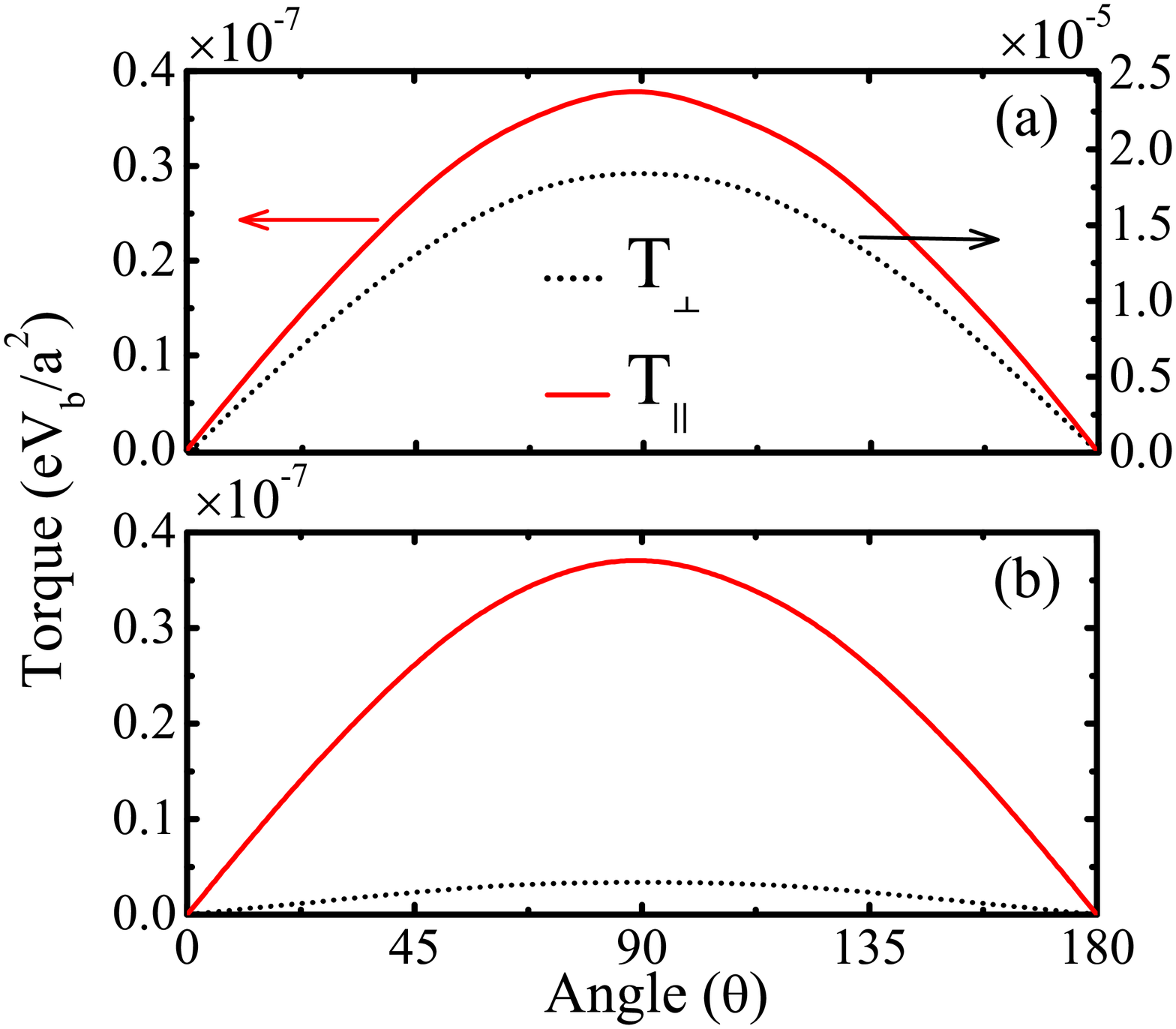,width=3.7in}}
\caption{The angular dependence of the in-plane and perpendicular components of the STT vector in {\em asymmetric} F$^\prime$/I/F MTJs illustrated in Fig.~1(a)
computed at zero temperature and for linear-response bias voltage $V_b$ using: (a) the proper gauge-invariant expression Eq.~\eqref{eq:rhoneqfullzero} for $\hat{\rho}_\mathrm{neq}$; and (b) the improper gauge-noninvariant expression Eq.~\eqref{eq:incorrect} for $\hat{\rho}_\mathrm{neq}$.}
\end{figurehere}

\subsection{Application to STT in semi-MTJs with strong interfacial Rashba SOC}\label{sec:semimtj}

To account for the Rashba SOC at the I/F interface in N/I/F semi-MTJ in Fig.~1(b), we add the following term
\begin{equation}\label{eq:sohamiltonian}
\varepsilon_{0,\mathbf{k}_\parallel} \mapsto \varepsilon_{0,\mathbf{k}_\parallel} + \alpha (\hat{\bm \sigma} \times \mathbf{k}_\parallel)  \cdot \mathbf{e}_x,
\end{equation}
to the on-site energy [see Hamiltonian in Eq.~\eqref{eq:fnhamiltonian}] of the first monolayer 0 of the F layer which is coupled to the I layer. The I layer has thickness $d_{\rm I}=5$ monolayers and $d_{\rm F}=20$ is the thickness of the free F layer. The parameter  \mbox{$\gamma_\mathrm{SO}=\alpha/2a$} quantifies the strength of the Rashba SOC, which we set at \mbox{$\gamma_\mathrm{SO}=0.1$ eV}. The on-site energy within the I layer is $\varepsilon_n = 6.0$ eV. The Fermi energy of this device in equilibrium is set at \mbox{$E_F=3.1$} eV.

The standard experiments to detect the presence of Rashba SOC at the I/F interfaces of N/I/F junctions involve measurements\cite{Gould2004,Moser2007,Wimmer2009,Chantis2007,Gmitra2009} of the so-called out-of-plane TAMR coefficient,\cite{Matos-Abiague2009,Mahfouzi2012} $\mathrm{TAMR}\,(\theta)=[R(\theta) - R(0)]/R(0)$. Here $R(0)$ is the resistance of semi-MTJ in Fig.~1(b) when the magnetization of its single F layer is parallel to the \mbox{$x$-axis} [i.e., direction of transport in Fig.~1(b)], and $R(\theta)$ is the junction resistance when magnetization is rotated by an angle $\theta$ with respect to the \mbox{$x$-axis} within the $xz$-plane. Since the interfacial SOC is linear in momentum, TAMR vanishes at the first order in  $\gamma_{\rm SO}$ after averaging over the Fermi sphere. However the F layer contains local exchange field and a net transfer of angular momentum occurs at the second order, so that TAMR $\propto \gamma_{\rm SO}^2$. This is also the origin of recently predicted\cite{Manchon2011,Mahfouzi2012a} unconventional TAMR-related STT in N/I/F  semi-MTJs in Fig.~1(b).

The semi-MTJs lack the spin-polarizing F$^\prime$ layer of conventional MTJs in Fig.~1(a), whose magnetization $\mathbf{m}^\prime$ together with the free magnetization $\mathbf{m}$ define the plane with respect to which STT vector is decomposed. Nevertheless, the usual torque components can be defined\cite{Manchon2011,Mahfouzi2012a} using
\begin{equation}\label{eq:sttcomponentsrashba}
{\bf T} = \mathbf{T}_{\parallel} + \mathbf{T}_{\perp} = \tau_{\parallel} {\bf m} \times ({\bf m} \times {\bf e}_x) + \tau_{\perp} {\bf m} \times {\bf e}_x.
\end{equation}
Here the direction of transport (${\bf e}_x$ in Fig.~1) replaces $\mathbf{m}^\prime$.

Since semi-MTJs in Fig.~1(b) are always structurally asymmetric, ${\bf T}_\perp  \propto V_b$ torque component is necessarily non-zero. To get its correct value requires to remove the equilibrium contribution ${\bf T}_\perp(V_b=0)$, which is accomplished by using the proper gauge-invariant $\hat{\rho}_\mathrm{neq}$ in Eq.~\eqref{eq:rhoneqfullzero} with the result shown in Fig.~3(a). Comparing this with ${\bf T}_\perp$ in Fig.~3(b), computed using gauge-noninvariant $\hat{\rho}_\mathrm{neq}$ in Eq.~\eqref{eq:incorrect}, reveals an order of magnitude discrepancy.

\begin{figurehere}\label{fig:fig3}
\centerline{\psfig{file=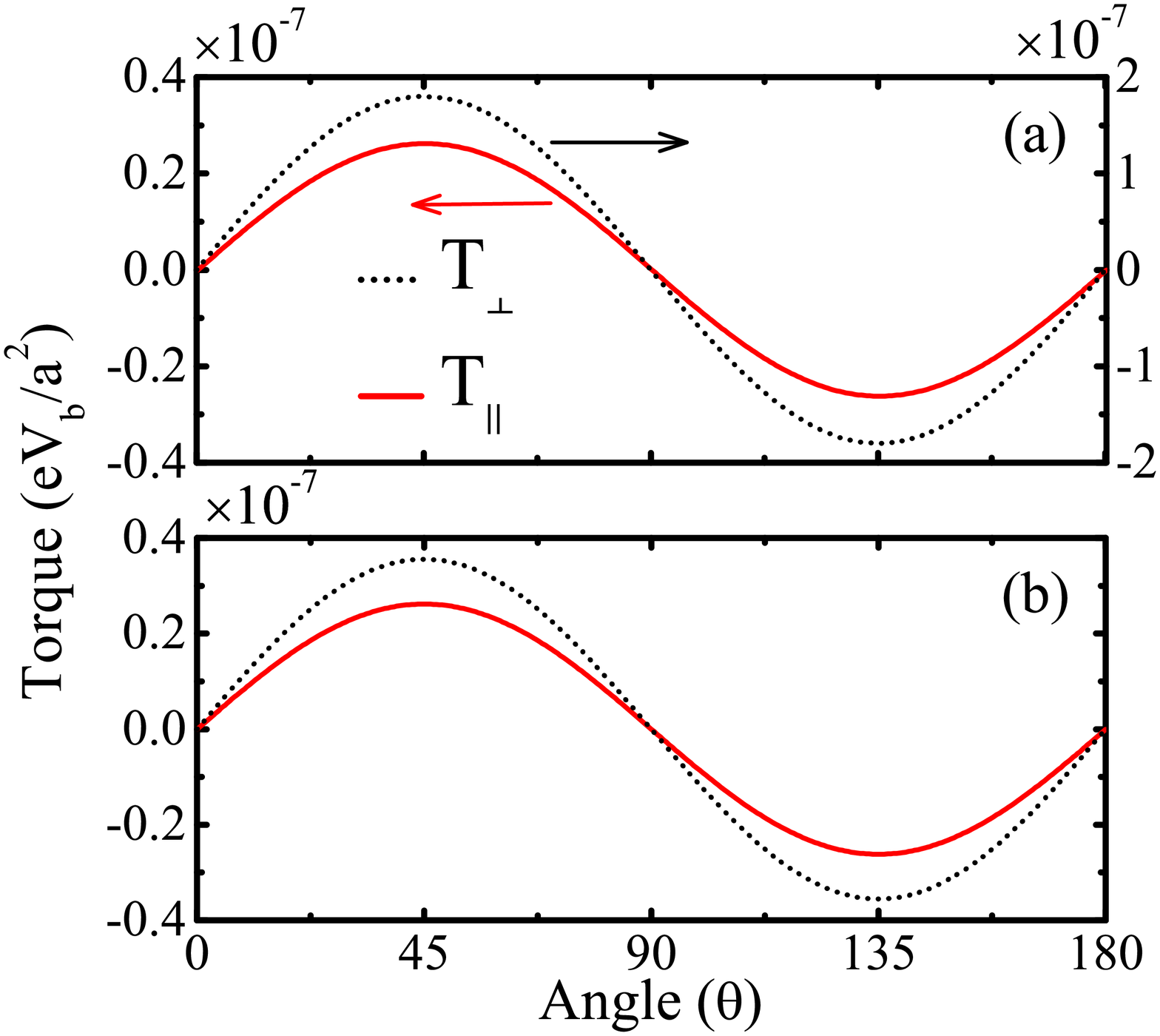,width=3.7in}}
\caption{The angular dependence of the in-plane and perpendicular components of the STT vector defined by Eq.~\eqref{eq:sttcomponentsrashba} for N/I/F semi-MTJs, illustrated in Fig.~1(b),  computed at zero temperature and for linear-response bias voltage $V_b$ using: (a) the proper gauge-invariant expression Eq.~\eqref{eq:rhoneqfullzero} for $\hat{\rho}_\mathrm{neq}$; and (b) the improper gauge-noninvariant expression Eq.~\eqref{eq:incorrect} for $\hat{\rho}_\mathrm{neq}$.}
\end{figurehere}

Unlike the symmetric angular dependence of conventional STT in MTJs discussed in Sec.~\ref{sec:mtj}, both ${\bf T}_\perp$ and ${\bf T}_\parallel$ exhibit $\propto \sin 2\theta$ angular dependence due to the existence of four stable magnetic states---two perpendicular to the I/F interface ($\theta=0,180^\circ$) and two parallel to that interface ($\theta = \pm 90^\circ$). The existence of both in-plane and perpendicular torque components in Eq.~\eqref{eq:sttcomponents} is analogous to  conventional STT in MTJs, where $\mathbf{T}_{\perp}$ in Eq.~\eqref{eq:sttcomponentsrashba} competes with the demagnetizing field and the perpendicular anisotropy while $\mathbf{T}_{\parallel}$ in Eq.~\eqref{eq:sttcomponentsrashba} competes with the damping. Thus, the unconventional TAMR-related STT can induce magnetization switching from out of plane to in plane and vice versa, as well as the current-driven magnetization precession.\cite{Manchon2011,Mahfouzi2012a,Manchon2011a}

\subsection{Application to SO torques in laterally patterned heterostructures with strong interfacial Rashba SOC} \label{sec:sotorque}

The early experimental confirmation\cite{Kato2004b,Wunderlich2005} of the spin Hall effect (SHE) in semiconductor devices with extrinsic (due to impurities) or intrinsic (due to band structure) SOC---where longitudinal unpolarized charge current generates  transverse pure spin current or spin accumulation at the lateral edges---has ignited theoretical studies of spin accumulation around the device edges in the diffusive\cite{Mishchenko2004,Nomura2005b,Adagideli2005} and ballistic\cite{Reynoso2006,Nikolic2005d,Zyuzin2007,Khaetskii2013} transport regimes. Since the prediction of the Edelstein effect,\cite{Edelstein1990} there has been also a lot of interest to understand how {\em nonequilibrium spin density} emerges in the interior of diffusive 2D systems with SOC within semiconductor heterostructures with structural inversion asymmetry.\cite{Inoue2003,Mishchenko2004} Most of such calculations have been focused on 2DEGs or 2D hole gases (2DHGs)  with the Rashba SOC (linear in momentum\cite{Reynoso2006,Mishchenko2004,Adagideli2005,Nikolic2005d,Zyuzin2007,Khaetskii2013} in the case of 2DEGs, or cubic\cite{Nomura2005b} in momentum in the case of 2DHGs) in different measurement geometries where 2D system is attached to two or more electrodes. For analytical calculations, it is advantageous to make one\cite{Mishchenko2004,Zyuzin2007,Khaetskii2013} or both\cite{Inoue2003} dimensions of a 2D system infinite. These studies have been typically conducted using either the Kubo formula\cite{Inoue2003,Nomura2005b} or the NEGF formalism.\cite{Reynoso2006,Mishchenko2004,Nikolic2005d}

The analogous problem arises in the analysis of SO torques in laterally patterned N/F/I heterostructures in Fig.~1(c), except that the effective mass Hamiltonian of a 2D system at the N/F interface
\begin{equation}\label{eq:rashba}
\hat{H} = \frac{\hat{\mathbf{p}}^2}{2m^*} + \frac{\alpha}{\hbar}(\hat{\bm \sigma} \times \hat{\mathbf{p}}) \cdot \mathbf{e}_z  - \frac{\Delta}{2} \mathbf{m} \cdot \hat{\bm \sigma},
\end{equation}
has both the Rashba (second) and the Zeeman (third) term. Here $\hat{\mathbf{p}} = (\hat{p}_x, \hat{p}_y)$ is the momentum operator and $\Delta$ is the mean-field exchange splitting due to the magnetization of the F layer pointing along the unit $\mathbf{m}$ vector. We recall that the same ferromagnetic Rashba Hamiltonian in Eq.~\eqref{eq:rashba}, with an additional term for static impurity potential, is also often used to study fundamental aspects of the anomalous Hall effect in itinerant metallic ferromagnets.\cite{Nagaosa2010}

While $\alpha$ in 2DEGs within typical semiconductor heterostructures is in the range 0.001--0.1 eV\AA{}, it can reach\cite{Gambardella2011} $\simeq 1$ eV\AA{} at the Pt/Co interface.\cite{Gambardella2011,Park2013} The very recent transport experiments\cite{Miron2010,Miron2011,Gambardella2011} have suggested that Rashba SOC could be responsible for the observed magnetization switching in a single F layer embedded between two asymmetric interfaces. For example, such effect was observed in Pt/Co/AlO$_x$ multilayers, but not in the inversion symmetric ones Pt/Co/Pt. The experiments reported in  Refs.~\refcite{Miron2010,Miron2011} have also utilized heavy atoms and surface oxidation to create strong out-of-plane potential gradient in Pt/Co/AlO$_x$ junctions and enhance the interfacial Rashba SOC.

When unpolarized current is injected into such 2DEG along the $x$-axis in Fig.~1(c), the ensuing nonequilibrium spin density can be obtained from
\begin{equation}\label{eq:spindensity}
\mathbf{S}_\mathrm{neq} = \frac{\hbar}{2} \mathrm{Tr}\, [\hat{\rho}_\mathrm{neq} \hat{\bm \sigma}].
\end{equation}
If the trace here is taken over the spin Hilbert space $\mathcal{H}_S$, one obtains the local spin density $\mathbf{S}_\mathrm{neq}(\mathbf{r})$, while taking the trace over the full Hilbert space $\mathcal{H}_\mathrm{O} \otimes \mathcal{H}_\mathrm{S}$ (where $\mathcal{H}_\mathrm{O}$ is the orbital space) gives total spin $\int d\mathbf{r} \, \mathbf{S}_\mathrm{neq}(\mathbf{r})$ [or $\sum_\mathbf{r} \mathbf{S}_\mathrm{neq}(\mathbf{r})$ in some discrete representation]. The knowledge of $\mathbf{S}_\mathrm{neq}$ makes it possible to compute\cite{Gambardella2011,Manchon2008,Manchon2009,Garate2009} the SO torque per unit volume acting on the magnetization of the F layer, $\mathbf{T}_\mathrm{SO} = \Delta (\mathbf{m} \times \mathbf{S}_\mathrm{neq}) /2$. The SO torque component along the $\mathbf{m} \times (\mathbf{j} \times \mathbf{e}_z)$ direction, where $\mathbf{j}$ is the in-plane current density, is field-like torque (i.e., analogous to $\mathbf{T}_\perp$ in MTJs discussed in Sec.~\ref{sec:mtj}) because it has the same form as precessional torque around an effective field in the $-\mathbf{j} \times \mathbf{e}_z$ direction. The other component along $\mathbf{m} \times [\mathbf{m} \times (\mathbf{j} \times \mathbf{e}_z)]$ direction has the same form as the damping-like torque toward a field in that same direction.

Below we use \mbox{$\mathbf{m} \equiv \mathbf{e}_z$} as in the experiments.\cite{Miron2010,Miron2011,Gambardella2011} Although realistic N/F interfaces are disordered,\cite{Wang2012,Haney2013} to illustrate yet another application of $\hat{\rho}_\mathrm{neq}$ in Eq.~\eqref{eq:rhoneqfullzero} we choose to analyze ballistic 2DEG for simplicity. Our 2DEG is infinite in the transverse (the $y$-axis in Fig.~1) direction while being  $L_x=100$ sites long in the $x$-direction of transport. Due to the periodicity in the $y$-direction, we perform an additional integration over $k_y$ to obtain
\begin{equation}
\mathbf{S}_\mathrm{neq}(x) = \frac{\hbar}{2}  \int \!\! dk_y \, \mathrm{Tr}\,[\hat{\rho}_\mathrm{neq}(k_y) \hat{\bm \sigma}].
\end{equation}

The discretization\cite{Nikolic2006} of Eq.~\eqref{eq:rashba} leads to a tight-binding like Hamiltonian defined on the square lattice of sites $\mathbf{n}=(n_x,n_y)$
\begin{eqnarray}\label{eq:tbh}
\hat{H}   =   \sum_{{\bf n},\sigma\sigma^\prime} \hat{c}_{{\bf
n}\sigma}^\dagger \left(\varepsilon_\mathbf{n} \delta_{\sigma\sigma'} - \frac{\Delta}{2} \mathbf{m} \cdot [\hat{\bm \sigma}]_{\sigma \sigma'} \right) \hat{c}_{{\bf n}\sigma^\prime} + \sum_{{\bf
nn'}\sigma\sigma'} \hat{c}_{{\bf n}\sigma}^\dagger t_{\bf
nn'}^{\sigma\sigma'}\hat{c}_{{\bf n'}\sigma'}.
\end{eqnarray}
whose nearest-neighbor hopping parameters are non-trivial $2 \times 2$ Hermitian  matrices  \mbox{${\bf t}_{\bf n'n}=({\bf t}_{\bf nn'})^\dagger$} in the spin space:
\begin{eqnarray}\label{eq:hopping}
{\bf t}_{\bf nn'}=\left\{
\begin{array}{cc}
-\gamma {\bf I}_{\rm s}-i\gamma_{\rm SO}\hat{\sigma}_y &
({\bf n}={\bf n}'+{\bf e}_x)\\
-\gamma {\bf I}_{\rm s}+i\gamma_{\rm SO}\hat{\sigma}_x &  ({\bf n}={\bf n}'+{\bf e}_y)
\end{array}\right.,
\end{eqnarray}
Here $\gamma=1.0$ eV is the orbital hopping, $\gamma_{\rm SO}=\alpha/2a$ is SO hopping which we set at $\gamma_\mathrm{SO}=0.1$ eV, and we chose $\Delta=0.6$ eV for the exchange splitting. The Fermi energy in this model is chosen close to the bottom of the band $E_F=-3.5$ eV in order to maintain  the parabolic energy-momentum dispersion of the original effective mass Hamiltonian Eq.~\eqref{eq:rashba}. While the on-site potential $\varepsilon_{\bf n}$ can be used to introduce disorder, we set $\varepsilon_{\bf n}=0$ in our ballistic 2DEG example.

In some prior studies of current-driven nonequilibrium spin density in 2DEGs with the Rashba SOC, as described by Eq.~\eqref{eq:tbh} with $\Delta=0$, one can find na\"{i}ve attempts to derive a linear-response formula for $\mathbf{S}_\mathrm{neq}(\mathbf{r})$ based on NEGFs. For example, Ref.~\refcite{Reynoso2006} starts from a general expression $\mathbf{S}_\mathrm{neq}(\mathbf{r})=\frac{\hbar}{4\pi i} \int dE\, \mathrm{Tr}_\mathrm{S}[\hat{\bm \sigma} \hat{G}^<(E)]$, obtained by combining Eqs.~\eqref{eq:expectation} and ~\eqref{eq:rhoglesser}, and then expands $\hat{G}^<(E)$ to linear order in the small bias voltage
\begin{eqnarray}
\hat{G}^<(E)  =  \hat{G}^<(E)\biggr|_{V_b=0}  - i\frac{eV_b}{2} \frac{\partial f(E)}{\partial E} [\hat{G}^r(E) (\hat{\Gamma}_L - \hat{\Gamma}_R) \hat{G}^a] + \mathcal{O}(V_b^2),
\end{eqnarray}
to arrive at the following formula
\begin{equation}
\mathbf{S}_\mathrm{neq}(\mathbf{r}) = \frac{\hbar e V_b}{4} \mathrm{Tr}_{\rm S} [\hat{\sigma}\{ \hat{G}^r(E_F) (\hat{\Gamma}_L - \hat{\Gamma}_R) \hat{G}^a(E_F)].
\end{equation}
This derivation assumes that the bias voltage is split using $V_L = -eV_b/2$ and $V_R = V_b/2$. However, such expression is {\em not} gauge-invariant since using $V_L = -V_b$ and $V_R = 0$ (or, equivalently, shifting the potential everywhere by a constant $-V_b/2$) would give different expansion
\begin{eqnarray}\label{eq:glessexpand}
\hat{G}^<(E)  =   \hat{G}^<(E)\biggr|_{V_b=0}  - i eV_b \frac{\partial f(E)}{\partial E} [\hat{G}^r(E) \hat{\Gamma}_L \hat{G}^a] + \mathcal{O}(V_b^2),
\end{eqnarray}
and different corresponding formula for $\mathbf{S}_\mathrm{neq}(\mathbf{r})$.

\begin{figurehere}\label{fig:fig4}
\centerline{\psfig{file=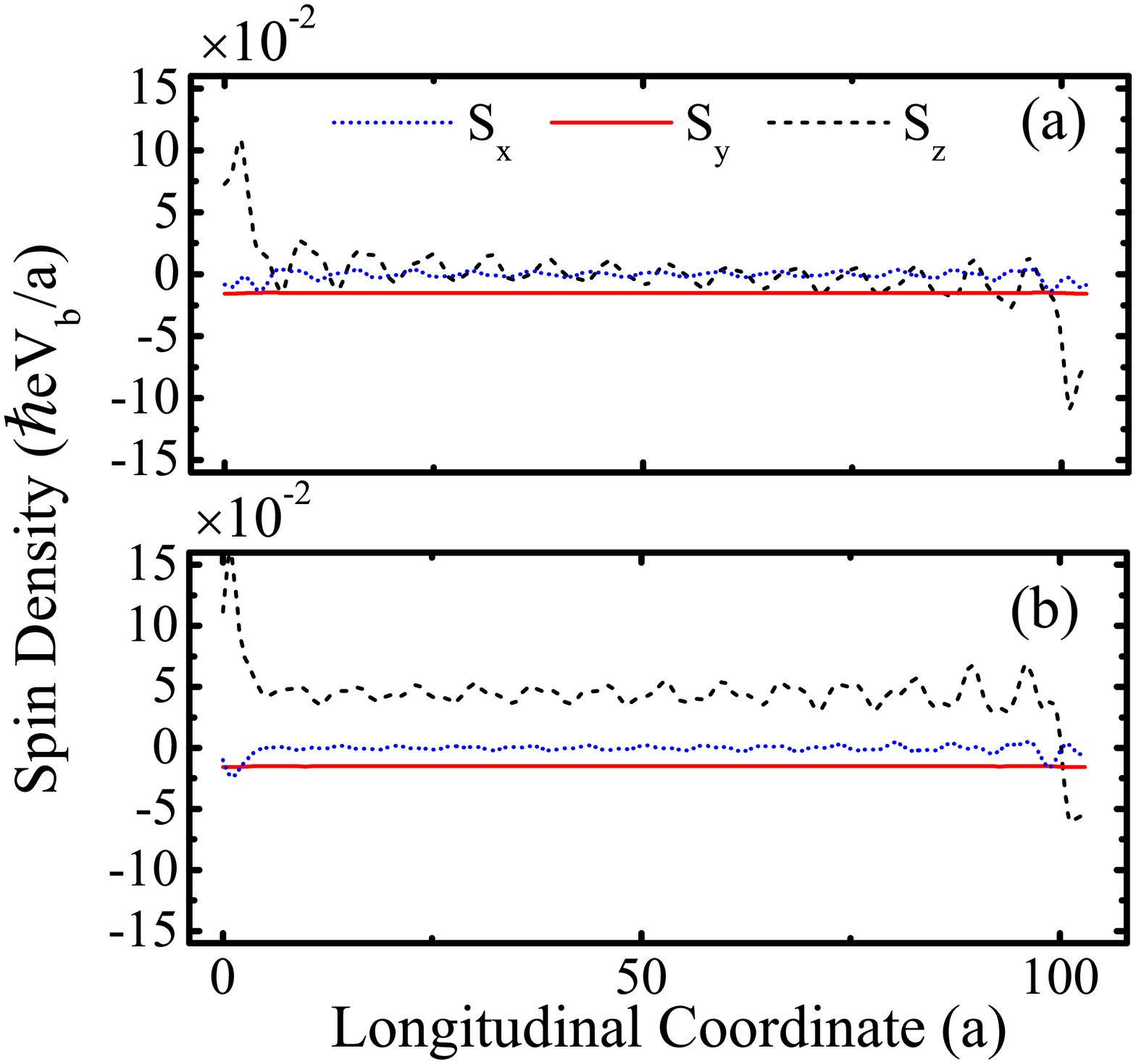,width=3.7in}}
\caption{The current-driven nonequilibrium spin density $\mathbf{S}_\mathrm{neq}(x)$ in the ferromagnetic 2DEG with the Rashba SOC, described by the Hamiltonian Eq.~\eqref{eq:rashba} with $\Delta \neq 0$, which is infinite in the $y$-direction and of finite-size in the transport $x$-direction. This quantity is computed at zero temperature and for the linear-response bias voltage $V_b$ using: (a) the proper gauge-invariant expression Eq.~\eqref{eq:rhoneqfullzero} for $\hat{\rho}_\mathrm{neq}$; and (b) the improper gauge-noninvariant expression Eq.~\eqref{eq:incorrect} for $\hat{\rho}_\mathrm{neq}$.}
\end{figurehere}

The usage of gauge-noninvariant expressions [note that Eq.~\eqref{eq:glessexpand} is equivalent to employing Eq.~\eqref{eq:incorrect}] does not affect previous results\cite{Nikolic2006,Reynoso2006,Nikolic2005d} obtained for the Rashba spin-split 2DEG in the absence of magnetization or external magnetic field,  where equilibrium spin density is absent due to the fact that SOC alone does not break time-reversal invariance. However, it  will lead to ambiguous results if applied to a 2D system described by the Hamiltonian in Eq.~\eqref{eq:rashba}, as demonstrated by Fig.~4. The difference between the results computed using improper Eq.~\eqref{eq:incorrect} and proper Eq.~\eqref{eq:rhoneqfullzero} expressions for $\hat{\rho}_\mathrm{neq}$ is less dramatic than in the case of $\mathbf{T}_\perp$ discussed in Secs.~\ref{sec:mtj} and ~\ref{sec:semimtj}.

\begin{figurehere}\label{fig:fig5}
\centerline{\psfig{file=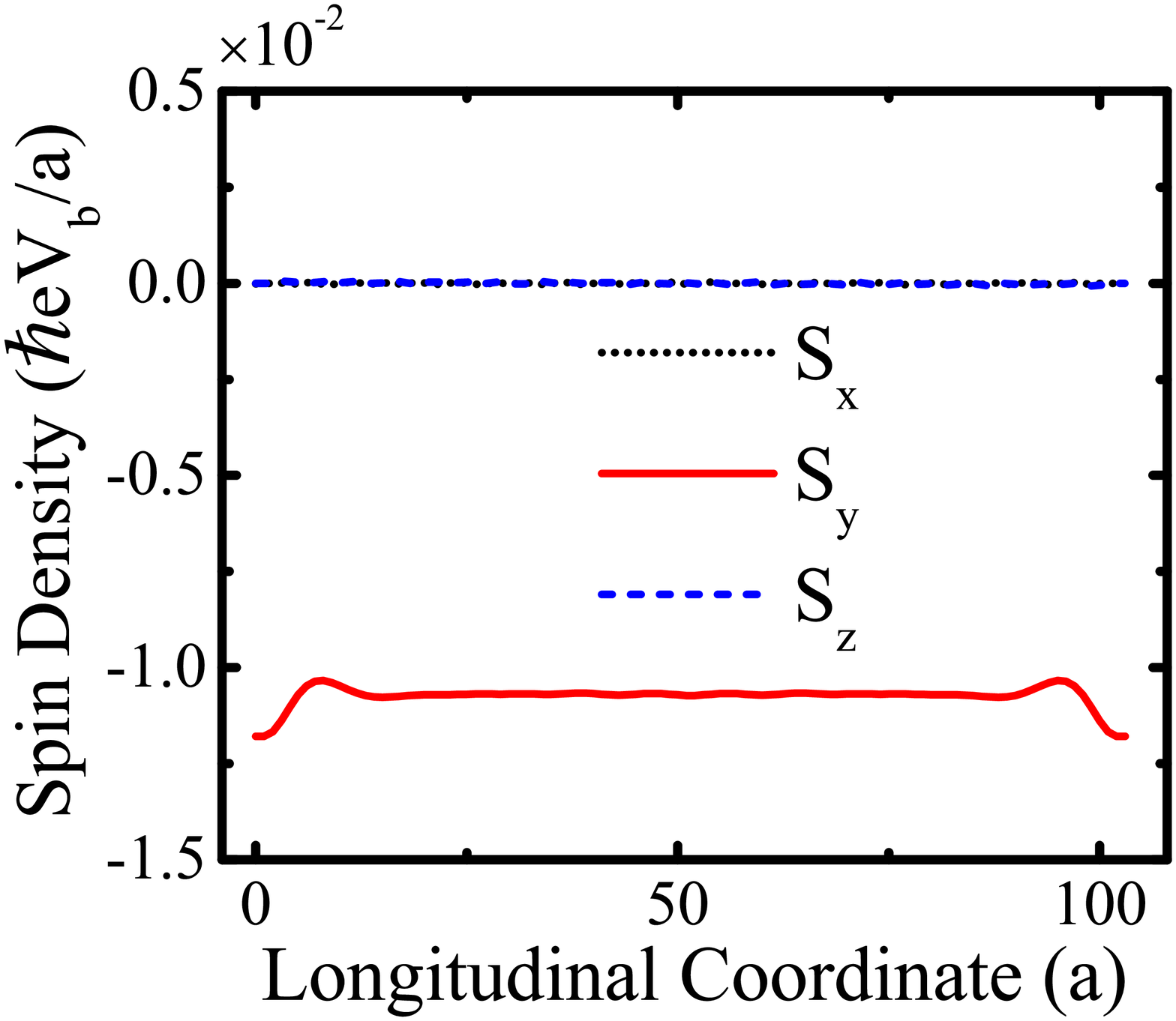,width=3.7in}}
\caption{The current-driven nonequilibrium spin density $\mathbf{S}_\mathrm{neq}(x)$ in the 2DEG with the Rashba SOC, described by the Hamiltonian Eq.~\eqref{eq:rashba} with $\Delta =0$, which is infinite in the $y$-direction and of finite-size in the transport $x$-direction. This quantity is computed at zero temperature and for the linear-response bias voltage $V_b$.}
\end{figurehere}

For comparison, we also plot $\mathbf{S}_\mathrm{neq}(x)$ in Fig.~5 for the case $\Delta=0$. This system exhibits large transverse nonequilibrium spin density $S_y^\mathrm{neq}(x) \neq 0$ [while $S_x^\mathrm{neq}(x)= S_z^\mathrm{neq}(x)=0$], akin to the Edelstein effect studied in infinite homogeneous diffusive 2DEGs.\cite{Edelstein1990,Inoue2003,Mishchenko2004} Introduction of non-zero magnetization into the Rashba Hamiltonian Eq.~\eqref{eq:rashba} leads to smaller $S_y^\mathrm{neq}(x)$, while also generating non-zero $S_x^\mathrm{neq}(x)$ and $S_z^\mathrm{neq}(x)$ with oscillatory spatial dependence in
Fig.~4(a) which is made possible by the ballistic nature of transport between the contacts. Thus, $\mathbf{T}_\mathrm{SO}$ in laterally patterned N/F/I heterostructure with clean N/F interface will be dominated by the field-like torque term along $\mathbf{m} \times \mathbf{e}_y$ direction since spatial integration of $S_x^\mathrm{neq}(x)$, which gives rise to damping-like torque component, averages to zero.

We note that recent experiments detecting current-driven magnetization switching in laterally patterned N/F/I heterostructures can be
interpreted using two different mechanisms: ({\em i}) current-induced $\mathbf{S}_\mathrm{neq}(\mathbf{r})$ at the N/F
interface due to strong Rashba SOC located at such interface;\cite{Miron2010,Miron2011,Gambardella2011,Suzuki2011} or ({\em ii}) SH current generated
within the bulk of a heavy metal N layer, which then flows perpendicularly through the N/F interface to generate STT on the F layer
magnetization.\cite{Liu2011b,Liu2012,Liu2012c} The latter case of SH torque is equivalent to a conventional torque in MTJs studied in Sec.~\ref{sec:mtj}
arising from a polarizing layer that would be located below the F layer with its magnetization pointing along the $y$-axis and the
current injection along the $z$-axis in Fig.~1(c). Its magnitude is largely determined by the bulk SH angle\cite{Liu2011b,Liu2012,Liu2012c} of the N layer, as well as by spin-dependent conductances of the N/F interface. Since Edelstein effect produces transverse nonequilibrium spin density, SO torque would act
predominantly as field-like torque,\cite{Miron2010,Miron2011,Gambardella2011,Suzuki2011} while the SHE mechanism
would generate dominant damping-like torque.\cite{Liu2011b,Liu2012,Liu2012c}

If both SH and SO torque mechanisms operate concurrently, the resulting torque is expected to be approximately the sum of the torques found in
the systems with one effect or the other.\cite{Haney2013} However, various experiments have reported conflicting results for the size and
direction of the torque. Furthermore, experimentally observed large sensitivity of torque on the thickness (even when changed by a
single atomic layer) of the F layer is presently difficult  to reconcile with predictions\cite{Manchon2012,Haney2013} based on either
of these mechanisms computed for simplistic model Hamiltonians.

Thus, first-principles studies investigating how structural, electronic and magnetic properties of N/F bilayers depend on the thickness of the layers are called for. For example, the very recent first-principles calculations have found strong Rashba SOC at the Pt/Co interfaces, but with higher order $k$-dependence and change in magnitude and sign from band to band.\cite{Park2013} The more sophisticated tight-binding Hamiltonians, fitted to first-principles computed electronic structure of bilayers, can then be combined with the proper gauge-invariant nonequilibrium density matrix in Eq.~\eqref{eq:rhoneqfull} to study 2D interfacial transport (to capture SO torques) or three-dimensional transport  (to capture SO and SH torques on the same footing) in arbitrary device geometry.

Experimentally, the key issue is to confirm the presence of strong interfacial SOC, which could be done using TAMR measurements\cite{Gould2004,Moser2007,Wimmer2009,Chantis2007,Gmitra2009} or the fact that SH torque itself would be modified when spin Hall current from the bulk of the N layer is traversing perpendicularly N/F interface supporting strong SOC---the angular dependence of SH torque would then exhibit additive combination\cite{Mahfouzi2012a} of conventional $\propto \sin \theta$ dependence shown in Fig.~2 and unconventional $\propto \sin 2\theta$ one shown in Fig.~3.

\section{Concluding remarks}~\label{sec:conclusions}

In conclusion, within the framework of nonequilibrium Green functions, we showed how to construct the proper {\em gauge-invariant} (i.e., independent of the reference level for electric potential) density matrix in steady-state nonequilibrium for an active region attached to two macroscopic reservoirs whose small electrochemical potential difference drives linear-response dc current (in the absence of inelastic processes in the active region). Our central expression---Eq.~\eqref{eq:rhoneqfull} or Eq.~\eqref{eq:rhoneqfullzero} at non-zero or zero temperature, respectively---contains three terms. One of
those terms is familiar from the usual two-terminal Landauer-type conductance formula as the expectation value of the total charge current in the electrodes, which is always a purely nonequilibrium quantity. The two additional terms ensure that any non-zero equilibrium expectation value of a physical quantity is properly removed from the formalism in gauge invariant fashion.

We illustrate the usage of the proper nonequilibrium density matrix by computing the field-like, which is non-zero even in equilibrium, and damping-like  components of the conventional torque in F$^\prime$/I/F MTJs or unconventional torque in N/I/F semi-MTJs with strong Rashba SOC at the I/F interface. The third application evaluates  current-driven nonequilibrium spin density in the ferromagnetic Rashba model, which yields the SO torque as one of the possible mechanisms behind magnetization switching of a single F layer recently observed\cite{Miron2010,Miron2011,Gambardella2011,Liu2011b,Liu2012,Liu2012c,Suzuki2011,Fan2013}  in laterally patterned N/F/I heterostructures with charge current injected parallel to the N/F plane. We compare these results with those obtained using the gauge-noninvariant expressions for the nonequilibrium density matrix employed in quantum transport literature to show how they lead to ambiguous values of current-driven field-like torque or nonequilibrium spin density due to improper removal of the corresponding equilibrium expectation values.

\nonumsection{Acknowledgments} \noindent We thank P. M. Haney, C. H. Lewenkopf, N. Nagaosa and I. Rungger for immensely valuable discussions and comments on several drafts. This material is based upon work supported by the US National Science Foundation under Grant No. ECCS 1202069.


\end{document}